\documentstyle[12pt,aaspp4,flushrt]{article}
\def\etal{\it et al. \rm }
\begin{document}

\title{The Butcher-Oemler Effect in Abell 2317}

\author{Karl D. Rakos\altaffilmark{1,2}}
\affil{Institute for Astronomy, University of Vienna, A-1180, Wein, Austria;
rakosch@astro1.ast.univie.ac.at}
\altaffiltext{1}{Visiting astronomer, Kitt Peak National Observatory,
National Optical Astronomy Observatories, which is operated by the
Association of Universities for Research in Astronomy, Inc. (AURA) under
cooperative agreement with the National Science Foundation.}
\altaffiltext{2}{Visiting Astronomer, Steward Observatory, University of
Arizona}

\author{Andrew P. Odell\altaffilmark{2}}
\affil{Department of Physics and Astronomy, Northern Arizona University, Box 6010,
Flagstaff, AZ 86011; andy.odell@nau.edu}

\author{James M. Schombert\altaffilmark{1}}
\affil{Department of Physics, University of Oregon, Eugene, OR 97403;
js@abyss.uoregon.edu}

\begin{abstract} 

This paper presents deep narrow band photometry of the cluster A2317
($z=0.211$) carried out using KPNO 4 m and Steward 2.3 m telescopes. Using
rest frame Str\"omgren photometry, it is determined that A2317 has an
unusually high fraction of blue galaxies (the Butcher-Oemler effect) for
its redshift ($f_B=0.35$).  We demonstrate that the ratio of blue to red
galaxies has a strong dependence on absolute magnitude such that blue
galaxies dominate the top of the luminosity function.  Spectrophotometric
classification shows that a majority of the red galaxies are E/S0's, with
a small number of reddened starburst galaxies.  Butcher-Oemler galaxies
are shown to be galaxies with star formation rates typical of late-type
spirals and irregular.  Starburst systems were typically found to be on
the lower end of the cluster luminosity function.  In addition, blue
galaxies are preferentially found in the outer edges of the cluster,
whereas the red galaxies are concentrated in the cluster core.

\end{abstract}

\section{INTRODUCTION}

Galaxy evolution has been a rapidly changing field in the last decade due
primarily to the explosion of observational information from the Hubble
Space Telescope and new generation of 8m+ class ground-based telescopes
(see Ellis 1997 for a review).  Early expectations from color evolution
models were that strong color evolution would only be visible in galaxies
beyond $z=1$ and, thus, not within the realm of optical telescopes due to
redshift effects.  However, two very surprising facts arose in the
mid-1980's; 1) the discovery of the Butcher-Oemler effect, the sharp rise
in the fraction of blue cluster galaxies at intermediate redshifts
($z=0.2$ to 0.4, Butcher and Oemler 1984), and 2) the excess of low
luminosity blue galaxies in the field population at intermediate redshifts
(Koo 1986, Tyson 1988).   These two discoveries indicated that galaxy
evolution was more than a simple calculation of color changes from a
passively evolving stellar population.  There are also significant changes
in the star formation history resulting in global changes to the cluster
luminosity function and population.

In an earlier paper of this series (Rakos and Schombert 1995, hereafter
RS95), we used a rest-frame, four-color photometry system (the Str\"omgren
system) to explore the color evolution of 17 clusters of galaxies from
redshifts of 0.2 to 0.9.  In that paper, the red population was shown to
follow closely the passive evolution models for a single burst population
with an epoch of galaxy formation near $z=5$. On the other hand, the blue
galaxies (the Butcher-Oemler population) showed a dramatic increase from
20\% at $z=0.4$ to 80\% at $z=0.9$, an even more rapid trend than an
extrapolation from intermediate redshift clusters would have predicted.
RS95 interpreted this trend as the rapid change of star-forming disk
galaxies into quiescent S0's or low surface brightness (LSB) galaxies.

The rapid rise of the blue galaxies in the Butcher-Oemler effect indicates
that a majority of a clusters population is involved in some amount of
star formation at epochs less than $z=1$.  Since our results for the red
population (ellipticals) are consistent with a formation epoch of $z=5$,
then either all the non-ellipticals in clusters (S0's, spirals and
irregulars) are star-forming at $z=1$, which then slowly discontinue star
formation by gas depletion or stripping, or else many of the members of
the blue population simply cease to exist by the present epoch.  The next
stage in our research in resolving this dilemma is to determine the nature
of Butcher-Oemler galaxies.

Our previous observations focused on the brightest galaxies in our cluster
sample, due to limitations set by the amount of observing time and
aperture size.  To explore the behavior of the red E/S0's and blue
Butcher-Oemler populations in more detail, we decided to return to one of
the RS95 clusters with a high fraction of blue galaxies $f_B$ and image
the cluster members to greater magnitude depth.  To this end we have
selected an intermediate redshift cluster, Abell 2317, for study.  Our
goal is to apply our new narrow band, photometric classification system to
the blue and red populations in a Butcher-Oemler cluster to 1) illuminate
the types of galaxies involved in the Butcher-Oemler effect, 2) determine
the dominance of the Butcher-Oemler galaxies to the cluster luminosity
function and 3) examine the spatial extend of the blue and red
populations.  Values of $H_o=50$ km sec$^{-1}$ Mpc$^{-1}$ and $q_o=0$ were
used throughout this paper.

\section{OBSERVATIONS}

Abell 2317 is a high declination cluster (19$^{\rm h}$ 08.5$^{\rm m}$
$+$68$^{\circ}$ 59.0$^\prime$) with a redshift of 0.211.  The cluster is
Coma-like in its richness (richness class 3, Abell, Corwin and Olowin
1989) and compact with a Bautz-Morgan class of II.  The brightest cluster
member has an apparent magnitude of 17.3 AB, and the cluster has a
moderate Galactic extinction of 0.32.

The observing procedure and reduction used herein is similar to that
described in RS95.  The observations of A2317 were taken with the KPNO 4m
PFCCD Te1K (binned 2 by 2) in June 1991 and with the Steward Observatory
2.3m in July 1995. The scale of the 4m telescope was 0.96 arcsec/pixel and
the exposure time was 25 minutes per filter divided in 5 frames of 300s
for cosmic-ray subtraction and flattening. The observations on the Steward
2.3m were restricted to the UV filter mostly, to increase the S/N for
faint galaxies since the QE of the 4m CCD was relatively poor in the blue
spectral region. The total exposure time on the 2.3 m was 3 hours divided
in frames of 900s. 

Photometry using 32 kpc apertures was performed on the final, co-added
frames.  Objects were selected based on detection in all four filters at
the 3$\sigma$ level.  There is no evidence from the color-magnitude
diagrams that this introduced a color bias to the sample.  Incompleteness
is evident below $m_{5500}=20$, the faintest objects are $m_{5500}=21$.
The $yz$ magnitudes were calibrated to a 5500\AA\ luminosity using
spectrophotometric standards (see below). Typical errors were 0.02 mags
for the brightest cluster members to 0.08 mags for objects below
$m_{5500}=20$.

According to Abell, Corwin and Olowin (1989), the size of the cluster is
15.2mm by 32.6mm on the Palomar Sky Survey, which corresponds to
approximately 15 by 32 arcminutes.  Our observations cover only the inner
8 by 8 arcminutes as restricted by the CCD field of view. Four-color
photometry in our modified Str\"omgren system ($uz$, $vz$, $bz$, $yz$, see
RS95) was obtained for 200 objects. Each of these filters are
approximately 200\AA\ wide and were specially designed so that they are
``redshifted" to the cluster wavelength in order to maintain a rest frame
color system. This provides a photometric system of determining cluster
membership without the use of redshift information due to the unique shape
of any galaxy's spectra around the 4000/AA/ break.  Our method is fully
discussed in Rakos, Schombert and Kreidl (1991) and Fiala, Rakos and
Stockton (1986), including the negligible effects from the cluster
velocity dispersion.

Using our photometric selection criteria, 112 objects (56\%) were
determined to be cluster members based on the behavior of the $mz$ index,
a measure of the SED around the 4000\AA\ break.  The $mz$ indice has the
advantage of being relatively flat within $\pm$3000 km sec$^{-1}$ of the
cluster redshift with sharp falloffs outside this redshift range.  This
maximizes the confidence of the membership of detected objects, while
minimizing the contamination from nearby field galaxies.  The remaining
44\% of the observed galaxies belong to the foreground or background
contamination, which is in agreement with the expectations from the
luminosity function of field galaxies in this magnitude range (Donnelly
1994).

In addition to resolving cluster membership, our four color photometry
system can also be used to produce a photometric classification of galaxy
type.  This classification is based on the 4000\AA\ colors (i.e. star
formation/activity) and only maps into morphology as one would expect star
formation history to map into the Hubble sequence.  In Rakos, Maindl and
Schombert (1996, hereafter RMS96), 140 high resolution, high S/N spectra
of galaxies were selected from the literature and reduced into synthetic
colors in our $uz$, $vz$, $bz$, $yz$ system.  From these colors, it was
demonstrated that two color and $mz$-color diagrams could be used to
classify galaxies into 4 simple types; ellipticals/S0's,
spirals/irregulars, Seyferts and starburst galaxies in a purely
spectrophotometric manner (see Figure 4 of that paper).  Guided by the
synthetic color results of RMS96, we have defined our photometric
classifications in the following manner (see Figures 1 and 2).  Any galaxy
with a red ($vz-yz$) color (i.e. $vz-yz > 0.65$) and $mz$ index greater
than $-$0.16 was classified as an E/S0 (an old stellar population
object).  The boundary between red spirals (Sa's) and E/S0's was clearly
defined at ($vz-yz)=0.65$ in RMS96 and is adopted herein.  Any galaxy with
a $mz$ index less than $-$0.16 is classified as a starburst.  A galaxy
that lies within 0.05 mag of the normal star-forming galaxy sequence
defined by Figure 3 of RMS96 was classified as a spiral/irregular (a
galaxy with a normal star formation rate).  And, lastly, any galaxy which
lies 3$\sigma$ blueward of the two color sequence and has a red $mz$ index
(indicative of the presence of strong non-star-forming emission lines
typical of AGN activity) is assigned a Seyfert classification.

The regions with the greatest uncertainty in photometric classification
are the boundaries between red spirals and ellipticals/S0's, the upper
bound of starbursts and the lower bound on Seyfert classification.  The
overlap between red spirals and elliptical/S0's has a minimal impact on
our results since red spirals are primarily large bulge-to-disk, gas-poor
objects, similar to S0's themselves.  Many evolutionary scenarios depict
early-type spirals as evolving to S0's after gas depletion and subsequent
reddening of the underlying stellar population.  The separation between
starbursts and spirals/irregulars is mostly a measure of the degree of
global star formation.  For the purposes of this study, the classification
of starburst has been conservative--the larger the separation from the
spiral/irregular mid-line the greater reddening, not the strength of the
starburst.

In addition to the colors, the $yz$ filter values can be converted to
$m_{5500}$.  The Str\"omgren system was originally designed only as a
color system; however, the $yz$ filter is centered on 5500\AA\ and it is
possible to link the $yz$ flux to a photon magnitude (i.e. the AB79 system
of Oke and Gunn 1983) through the use of spectrophotometry standards.
This procedure was described in detail in Rakos, Fiala and Schombert
(1988) and is used to determine the $m_{5500}$ values for all the cluster
members.  The incompleteness of the sample is not serious until below
$m_{5500}=20$, which corresponds to $M_B=-$20.7 ($H_o=50$).  The number of
fainter galaxies is, in reality, much larger than in our sample, but our
results will only depend on the bright end of the luminosity function.  In
order to save space and publication charges, the data for the 112 cluster
members of A2317 can be obtained from zebu.uoregon.edu/$\sim$js. This
ASCII file contains pixel coordinates, $m_{5500}$, $uz,vz,bz,yz$ colors,
errors and our photometric classifications.

\section{DISCUSSION}
\subsection{Photometric Classification}

The extragalactic meaning of the Str\"omgren colors are described in
detail in our previous papers (see RS95).  Briefly, the Str\"omgren
colors, centered around the 4000\AA\ break, are crude estimators of recent
star formation, mean age and metallicity.  In previous papers, we have
used ($vz-yz$) as a measure of global metallicity and ($bz-yz$) as a
measure of mean age of the underlying stellar population.  However,
precise understanding of these colors requires comparison to SED models
with various assumptions (e.g. redshift of formation, mean metallicity and
IMF), since varying star formation histories inhibit a unique
interpretation of the colors.  Recent star formation can also strongly
influence ($vz-yz$), and in this paper we are expanding the use of our
Str\"omgren filters to explore the star formation history of
Butcher-Oemler galaxies.

One advantage of the Str\"omgren colors (or any 4000\AA\ color system) is
their strong dependence on recent star formation and their sensitivity to
deviations in star formation rates that signal starburst activity.  RMS96
outlines the $uvby$ system sensitive to star formation with the starburst
models of Lehnert and Heckman (1996, see Figure 6 of RMS96).  As discussed
in \S 2, we have classified the cluster members based on their colors and
$mz$ indices by a prescription developed in RMS96.  We have divided the
photometric classifications into four subtypes as current colors reflect
into star formation history; 1) elliptical/S0 (E/S0), 2) spiral/irregular
(Sp/Irr), 3) starburst and 4) Seyfert.  Operationally, this subtypes are
defined as 1) red, non-star-forming colors, 2) blue star forming colors as
defined by Hubble types Sa to Irr, 3) having deviant colors indicative of
strong star formation combined with reddening, and 4) peculiar fluxes in
the $vz$ filter that signal AGN phenomenon.  The relationships between
galaxy type and color indices is shown in Figures 1 and 2, the $mz$-color
and two color diagrams.

Starburst galaxies cover a range of colors from blue to red, but are
located on the reddened side of the two color diagram.  The analysis of
interacting and merging galaxies in RMS96 demonstrated that the typical
colors of a tidally induced starburst system lie in this portion of the
two color diagram (Figure 2).  It is possible to differentiate between
starburst galaxies shrouded in dust with strong reddening (IRAS
starbursts, Lehnert and Heckman 1996) and the strong ultraviolet radiation
of WR galaxies simply by the distance the galaxy lies from the Sp/Irr
sequence along the reddening line, although the exact amount of reddening
is subjective.  The reddening line in Figure 2 indicate that a majority of
the starburst galaxies have Sb or later colors with significant extinction
by dust.

The reader is cautioned that this classification scheme is based on the
integrated colors and is not a morphological system. The designation of
E/S0, Sp/Irr or starburst only refers to the current star formation rate
and recent SF history of the galaxy and not to its morphological
appearance, the existence of spiral arms or a bulge-to-disk ratio.  For
example, there is no discrimination between ellipticals and S0's since
their global colors are identical.  Many of the galaxies on the blue side
of the E/S0 boundary may, in fact, be large-bulge Sa's whose blue disk
light is overwhelmed by the more dominant red bulge. Galaxies on the red
side of the Sp/Irr sequence may be the new class of ``E+A'' post-starburst
galaxies whose smooth morphology is not measured or simply disk systems
with very low star formation rates.  However, there has always been a
strong relationship between galaxy color and morphology such that it can
be assumed that a majority of the Sp/Irr systems, classified by color, are
disk galaxies.  Likewise, the starburst systems are most likely
interacting or merging galaxies and Seyferts are disk galaxies with AGN
cores.

Judging from the results of the photometric classification, A2317 is
similar to other Butcher-Oemler clusters by being deficient in E/S0's and
overabundant in late-type galaxies.  The population fractions are listed
in Table 1 and, in particular, we find that 47 (42\%) of the cluster
members are E/S0's, whereas 29 (26\%) are Sp/Irr and 25 (22\%) are
starbursts.  The remaining 11 (10\%) are classified as Seyferts.  Typical
E:S0:Sp/Irr ratios for present-day clusters are 20\%/40\%/40\% (Oemler
1992), based on morphological classification of the nearby cluster
galaxies.  It is assumed that all the galaxies photometrically classified
galaxies as Sp/Irr, starburst and Seyfert would be contained in a
late-type category and, thus, A2317 is 20\% overabundant in late-type
galaxies as compared to present-day clusters.

Of the blue population, a majority (46\%) are Sp/Irr with Seyferts and
starbursts equally split for the remainder.  The red population is
primarily E/S0's (64\%) with 15\% classed as Sp/Irr and 21\% as highly
reddened starburst galaxies.  Ignoring the low luminosity starbursts, the
ratio of E/S0's to red Sp/Irr's is exactly the same as present-day
clusters (i.e. 4 to 1).  This leads us to speculate that the current
starburst population fades to develop into the faint dE/dI population that
so dominates present-day cluster counts. The ratio of the remaining bright
galaxies will then match the morphological fractions of present-day
clusters.

The large number of Seyferts in A2317 is anomalous for a rich cluster
environment.  The fraction of Seyferts found in local clusters is
extremely low, less than 1\% (Dressler and Shectman 1988) and the fraction
found in the local field is less than 2\% (Huchra and Burg 1992) compared
to our observed value of 11\% for A2317.  However, this value is similar
to the fraction of AGN galaxies in 3C 295 ($z=0.465$, $f_B=0.22$, Dressler
and Gunn 1983) and a recent study by Sarajedini \etal (1996) finds the
fraction of AGN's in the field to be 10\% between $z=0.2$ and 0.6. Thus,
our value of 10\% is an exact match with the sharp increase of AGN
activity at these redshifts in both cluster and field environments.

\subsection{Color-Magnitude Diagram}

The color-magnitude (C-M) diagrams for colors ($bz-yz$) and ($vz-yz$) are
shown in Figures 3 and 4.  Immediately obvious are the distinct C-M
relationships for ellipticals and S0's.  The C-M relation for a single
burst stellar population, such as ellipticals and S0's, is a reflection
of the mass-metallicity correlation predicted by simple closed-box models
of galaxy evolution (Faber 1973, Tinsley 1980).  The greater the mass of a
galaxy the higher its luminosity and its SN rate.  The higher SN rate
results in a more rapid initial enrichment of the first generation of
stars and, therefore, a redder red giant branch for the composite
population.  The $vz$ filter is located over several strong metal lines in
an old population spectrum and, thus, the ($vz-yz$) colors are more
sensitive to metallicity changes compared to ($bz-yz$).  This can be seen
in Figure 3, ($bz-yz$) versus $m_{5500}$ were the E/S0 sequence is
effectively independent of continuum color.  On the other hand, Figure 4,
the ($vz-yz$) versus $m_{5500}$ diagram, displays a clear trend of redder
metallicity color with higher luminosity (mass).

The solid lines in Figures 3 and 4 are the C-M relations from local
populations discussed in Schombert \etal (1993).  The ($bz-yz$) relation
exactly matches the A2317 galaxies, whereas the E/S0's appear to have a
sharper slope in the ($vz-yz$) diagram.  There is no direct reason for
believing that the slope of the C-M relation should change dramatically
with redshift.  There will certainly be changes in luminosity with the
galaxy population; for example Schade (1996) finds that the surface
brightness increases to 1.5 mags for an elliptical population at
$z=0.6$.  However, changes in luminosity of up to 0.5 mag in Figure 4
would be undetectable in the data since the correlation is nearly parallel
to apparent magnitude.  A steeper slope could suggest an age difference
between low and high luminosity E/S0's (Bender \etal 1996), however there
is no evidence of changes in the C-M slope in clusters at $z=0.4$ (Oemler,
Dressler and Butcher 1997).

The blue side of the C-M diagrams are populated with Sp/Irr's, Seyferts
and a few starburst systems.  Unlike other Butcher-Oemler clusters at this
redshift, a majority of the brightest galaxies are blue (see the next
section).  The Seyferts are scattered amongst the brightest Sp/Irr's;
however the starburst systems tend to be lower in luminosity than the
other galaxy types.  The starburst systems also spread across a range of
colors, as would be expected from their variable reddening.

\subsection{Butcher-Oemler effect}

The original broadband definition of the fraction of blue to red galaxies,
$f_B$, given by Butcher and Oemler (1984), is the ratio of galaxies 0.2
mag bluer than the mean color of the E/S0 sequence (after K-corrections)
to the total number of galaxies in the cluster.  As discussed in Rakos,
Schombert and Kreidl (1991), a sample of nearby spirals and irregulars is
used to determine a value of ($bz-yz$) (continuum color) in our filter
system that separates star-forming from quiescent galaxies. From this
previous analysis, we have defined the fraction of blue galaxies, $f_B$,
as the ratio of the number of galaxies bluer than ($bz-yz$)=0.22 to the
total number of galaxies.  Since the mean ($bz-yz$) color of a present-day
elliptical is 0.37 (Schombert \etal 1993) and ($bz-yz$) maps into ($B-V$)
in a linear fashion with a slope of 1.33 (Matsushima 1969), then a cut at
($bz-yz$)=0.22 is effectively the same as Butcher and Oemler's 0.2 mag
selection and the two values are directly comparable.  In fact, we have
argued that our narrow band system is superior to the broadband ($B-V$)
criteria due to our filters' unique locations on continuum regions of a
galaxy's SED.  The interpretation of the Butcher-Oemler effect within our
filter system is a comparison of the mean temperature of the composite
stellar photosphere, undistorted by emission lines or
foreground/background contamination.

Applying this criteria to A2317 produces a $f_B$ value of 0.35$\pm$0.06,
which for this redshift range makes it a cluster unusually rich in blue
galaxies (note that a typographical error in RS95 lists A2317 with a $f_B$
value of 0.51; the correct value is 0.31 for that study, well within the
errors of this current, deeper dataset).  Figure 3 shows a dashed line
for the ($bz-yz$) cutoff for red/blue classification.  Note that all the
E/S0 type galaxies from the photometric classification are within the red
boundary.  This does not imply that the post-starburst ``E+A'' type
galaxies are missing from A2317.  This type of galaxy is a mismatch of
morphology with color (a smooth morphology with a blue underlying stellar
population suggestive of recent star formation) and would have been
classed as an Sp/Irr in our system.  Nine Sp/Irr's also lie within the red
population region demonstrating that a simple color division of a cluster
population can overlook a more subtle variation in multicolor space. These
objects are most likely early-type spirals with large bulges that dominate
the colors, but with small star forming disks that are still detectable in
our multicolor space.  There is also the possibility that these galaxies
are young S0's with post-starburst disks (Bothun and Gregg 1990).
Numerous starburst class galaxies also lie within the red region, however
inspection of Figure 2 shows that these are all simply very reddened
systems.

Figures 3 and 4 also show that the brightest galaxies have bluest colors.
In fact, the measured $f_B$ value for the cluster is strongly dependent on
the cut-off magnitude of the sample.  This can be seen in Figure 5 which
displays $f_B$ as a function of $m_{5500}$.  The fraction of blue galaxies
is over 75\% in the highest magnitude bin dropping to only 10\% for
galaxies with luminosities between $L^*$ and 1/2 $L^*$ and then rising
again for galaxies with luminosities below 1/2 $L^*$.  The change in $f_B$
is significant and implies that the scatter in the Butcher-Oemler effect
with redshift (see Figure 4 of RS95), particularly at low redshifts, may
be due to limiting magnitude effects combined with evolutionary changes in
the luminosity function with redshift.  For example, the brightest blue
galaxies may have very short lifetimes, not in terms of existence as
self-gravitating entities, but in terms of visibility.  In RS95, a
scenario was proposed were Butcher-Oemler galaxies fade into LSB galaxies
or low luminosity dwarf galaxies after their initial burst of star
formation.  There is some evidence of this effect in other cluster studies
(see Schade 1996) in that blue galaxies in distant clusters display a
measured increase of 1 to 1.5 mags in mean surface brightness.

The rise in $f_B$ at the lowest luminosities in A2317 is suggestive of an
increasing contribution from a dwarf galaxy population.  Although this
luminosity range is not completely sampled by our study, the minimum in
$f_B$ at 19.5 (see Figure 5) is also at the same absolute luminosity where
in the Virgo cluster there is a shift in the luminosity function from
giant galaxies to a high fraction of dwarf Irr, Im and Sm's.  The class of
dE galaxies dominate the luminosity function at very low luminosities
(Sandage, Binggeli and Tammann 1985), but blue late-type systems at just
below $L^*$ are sufficiently numerous to produce a rise in the blue galaxy
fraction.  We note that most of the galaxies classified as starburst are
in the lower luminosity bins.  Although many of the starburst galaxies are
not blue galaxies due to reddening effects, the larger fraction below
$L^*$ would lend support to the proposal of Koo \etal (1997) that a
majority of the starburst systems in clusters at intermediate redshifts
are dwarf galaxies undergoing a phase of star formation.  These bursting
dwarfs would have a relatively high luminosity for a short timescale, then
fade to normal dwarf luminosities by the present epoch.

Although the morphological type of the brightest blue galaxies is not
known (our photometric classification does not distinguish between early
and late-type spirals), results from recent HST imaging of distant
clusters suggests that these systems are structurally late-type spirals
(i.e. small B/D ratios) with slightly increased levels of star formation
activity (Oemler, Dressler and Butcher 1997).  Along the same lines, the
galaxies photometrically classified as starburst can linked with the
merging/interacting systems found by the Oemler \etal (1997) study.  The
average merging/interacting systems in the four clusters from Oemler \etal
is 17\%, compared to the 22\% fraction of starburst galaxies in A2317, in
agreement with our expectation that the strongest star formation in this
Butcher-Oemler cluster is tidally induced (see below).  Regardless of the
appearance of either the Sp/Irr or the starburst galaxies, the main
conclusion from Figure 4 is in agreement with the morphological studies, 
that the blue Butcher-Oemler population is primarily composed of galaxies
with normal star formation rates typical of present-day spirals and
irregulars.

\subsection{Galaxy Harassment and the Origin of the Blue Population}

The spatial distribution of blue and red galaxies is shown in Figure 6, a
histogram of number density (normalized by projected area) as a function
of radius from the cluster core.  In this figure, the cluster core is
calculated from the geometric mean of the photometrically determined
cluster members.  The results below were insensitive to changes in the
center at the level of 20 arcsecs.  Immediately obvious from Figure 6 is
that the red galaxies have the highest densities at the cluster center
with a smooth density dropoff toward the outer part of the cluster.  In
contrast, the blue galaxies have a marked drop in their density for the
inner third of the cluster with a roughly uniform spatial density in the
outer part of the cluster.  There is no obvious trend between Sp/Irr,
Seyfert or starburst types.

The blue population has long been suspected of avoiding the densest
regions of a cluster core (see Dressler 1993), and a recent study by
Abraham \etal (1996) has shown a distinct drop in $f_B$ towards the core
of A2390, another Butcher-Oemler cluster.  Carlberg \etal (1997), in a
study of the CNOC cluster sample, demonstrate that not only is there a
drop in the blue galaxy fraction at cluster cores, but that the red
population has a smaller velocity dispersion by a factor of 1.3 compared
to the blue population.

The difference in the red and blue population dynamics has led the Abraham
\etal group to propose that the Butcher-Oemler effect is due to an
infalling field population.  The subsequent dying out of the blue
population is due to truncation of star formation from stripping by the
intracluster medium.  The infalling spirals then evolve into red S0's,
which is in agreement with the large fraction of S0's found in present-day
clusters.  Unfortunately, the HST morphological data does not agree with
this scenario in the sense that the blue population is shown to be
composed of small bulge, late-type spirals which will not fade into large
bulge S0's.  The argument made in Abraham \etal is that the relevant
parameter is the ratio of the bulge to disk luminosity, which will change
dramatically with the cessation of star formation in the disk.  Of course,
bulge to disk ratio is only one component of the morphological
classification of spirals (the other two components are tightness of
spiral arms and degree of resolution of the arms into stars), so the
question of whether the Butcher-Oemler population evolves from late-type
to early-type will remain unresolved until a detailed surface photometric
decomposition of the blue galaxies is made.

If Oemler \etal (1997) have correctly classified the blue galaxy
population as late-type spirals with small bulges, then the origin and
destiny of the Butcher-Oemler galaxies remains unclear.  The high fraction
of mergers/interactions is suggestive of tidally induced star formation,
yet this does not resolve the fate of the blue galaxies, since even the
abrupt cessation of star formation would still leave a population of red,
late-type spirals in present-day clusters, which is not seen.  Tidal
interactions may well explain the high fraction of starburst galaxies;
these would have their origin in gas-rich dwarf galaxies undergoing a
short, but intense, tidally induced starburst (Koo \etal 1997).  It should
be noted however, that the cluster environment is the most surprising
arena for the type of interactions that can lead to star formation
bursts.  The orbits of cluster galaxies are primarily radial and the
typical velocities are high.  This makes any encounter with another galaxy
shortlived, with little impulse being transferred as required to shock the
incumbent molecular clouds into a nuclear starburst.

Recently, a new mechanism for cluster induced star formation has been
proposed.  This method, called galaxy harassment (Moore \etal 1996),
emphasizes the influence of the cluster tidal field and rapid, impulse
encounters with massive galaxies.  These two processes conspire to not
only raise the luminosity of cluster spirals, but to increase their
visibility, hence detectability.  One of the predictions of galaxy
harassment is that galaxies in the cores of clusters will be older than
galaxies at the edges.  In terms of star formation history, this is
exactly what Figure 6 has demonstrated in A2317.  The blue population
(i.e. the harassed population) is primarily located in the outer 2/3's of
the cluster.  Other explanations of the blue population, such as ram
pressure stripping or infall of blue galaxy rich subclusters, would have
the blue population primarily confined to the cluster edges.  

\subsection{Fate of the Butcher-Oemler Population}

Regardless of the origin of the blue population, its fate is obvious.
These galaxies do not exist in present-day clusters and, therefore, must
either be destroyed or reduced to the luminosity of dwarf galaxies.  Two
points concerning A2317 make this fate difficult to reconcile with our
current understanding of galaxy evolution.  The first is that the
Butcher-Oemler population is numerous and the second is that this
population contains many bright members.  Despite any changes in total
luminosity or surface brightness, the blue population makes up a
significant fraction of a cluster numerically, and this value increases
with redshift.  Since our survey of present-day clusters is complete to
extremely low luminosity and surface brightness, this implies that the
blue population must eventually be eliminated as recognizable units or
transformed in to objects with much smaller luminosity.  This conclusion
is supported by the observation in Table 1 that the red population already
contains the correct morphological mix found in local clusters.  A
conversion of blue to red galaxies is not needed to resolve morphological
distributions.

One point to remember is that the destruction of the Butcher-Oemler
population is only a radical proposition if these galaxies contain a
significant fraction of the mass of the cluster rather than luminosity.
For example, the blue population may be a specific population of objects
with very low $M/L$'s or extremely short lifetimes.  Some of the blue
population may be bursting dwarf galaxies, as proposed by Koo \etal
(1997), and as is suggested by the population of low luminosity starbursts
in this study; however, the HST imaging results indicate that most of the
blue population, particularly the bright objects, are normal late-type
spirals by their morphological appearance.

Interestingly enough, there already exists a class of galaxies which
combines late-type morphological appearance with a particular frailty to
tidal forces.  These are the LSB galaxies of recent attention in field
surveys and dark matter investigations (see Impey and Bothun 1997 for a
review).  Recent work has shown that LSB galaxies, which are by definition
low in luminosity density, are gas-rich and low in mass density (de Blok and
McGaugh 1996).  Thus, LSB galaxies would be particularly susceptible to
tidally induced star formation (Mihos, McGaugh and de Blok 1997). LSB
galaxies also have an unusually high incidence of low luminosity AGN activity
(Impey and Bothun 1997), which may account for the large number of
Seyferts in our sample as starburst gas is also dumped onto a central
engine.  

The scenario proposed here is similar to the one we proposed in RS95, that
the Butcher-Oemler population is an evolved set of LSB galaxies.  The
mechanism of galaxy harassment or ram pressure induces highly efficient
star formation in these dark matter systems that increases their
visibility and actual luminosity.  Later encounters with the cluster core
destroys this low mass density system by tidal forces.  Thus, we believe
that the Butcher-Oemler cluster will evolve into future cD clusters from
the released stellar material of the disrupted LSB population.

\section{SUMMARY}

This paper has presented deep multi-color photometry of the intermediate
redshift cluster A2317.  The color system used herein is a specific form
of the Str\"omgren color system, matched to the rest frame of A2317, in
order to understand the components of the blue, Butcher-Oemler population
known to inhabit this cluster.  Our primary results are summarized as:

\noindent (1) 112 cluster members were identified and classified by our
photometric system with four subtypes; 1) E/S0, 2) Sp/Irr, 3) Seyfert or
4) starburst.  The fractional population is listed in Table 1 and color
diagrams are shown in Figures 1 and 2.  The fraction of blue galaxies is
$f_B$=0.35.

\noindent (2) The C-M diagrams in Figures 3 and 4 show that the blue
population dominates the top and bottom of the luminosity function.  That
is to say, the brightest galaxies are blue, with a decrease at
intermediate magnitudes, then a rise at low luminosities (see Figure 5).
There is a steepening of the C-M relation for E/S0's as compared to the
locally determined relation, but the difference is not statistically
significant within the context of only one cluster determination.

\noindent (3) The Butcher-Oemler population is, based on our photometric
classification system, primarily a population of galaxies with normal star
formation rates.  Although there are anomalously high fractions of
Seyferts and starburst galaxies in A2317, most of the blue galaxies are
classified as Sp/Irr.  The starburst systems tend to be of low luminosity
suggesting they may be bursting dwarfs (Koo \etal 1997).  The above
photometric classification of the Butcher-Oemler population, based on
their star formation history, is in agreement with their HST image
classifications as late-type spirals (Oemler \etal 1997).

\noindent (4) The spatial location of the blue population (see Figure 6)
indicates that the Butcher-Oemler effect is confined to the outer cluster
regions with the red population dominating the cluster core.  This result
is in agreement with the prediction of galaxy harassment models, where the
cluster tidal field and impulse encounters with more massive galaxies
results in tidally induced star formation or infall models of spiral-rich,
merging subclusters.

The fate of the Butcher-Oemler population is still undetermined since they
are too numerous and too recent to simply fade from view by the present
epoch.  We renew our arguments that the most plausible fate for the blue
population is destruction, and that the Butcher-Oemler population is
composed of low density LSB galaxies that are particularly susceptible to
disruption in the dense cluster cores.

\acknowledgements
The authors wish to thank the director and staff of KPNO and Steward
Observatory for granting time for this project.  Financial support from the
Austrian Fonds zur Foerderung der Wissenschaftlichen Forschung is
gratefully acknowledged.  This research has made use of the NASA/IPAC
Extragalactic Database (NED) which is operated by the Jet Propulsion
Laboratory, California Institute of Technology, under contract with the
National Aeronautics and Space Administration.

\clearpage

\figcaption{The $mz$ index versus metallicity color ($vz-yz$).  Photometric
classifications are shown.  All starburst systems are defined by $mz$
indices less than $-$0.16.}

\figcaption{Two color diagram for A2317 members.  Photometric
classifications are shown.  Seyfert galaxies are defined by being bluer
than more than 3$\sigma$ from the Sp/Irr sequence. A reddening vector of 1
mag is shown.  Note that all starburst systems display from 1 to 3 mags of
reddening, most probably an IRAS style of starburst shrouded in dust.}

\figcaption{Color-Magnitude diagram for the continuum color, ($bz-yz$).  An
$L^*$ galaxy has an apparent mag of 19.5 ($H_o=50$).  Photometric
classifications are shown.  The dotted line is the cutoff of $bz-yz=0.22$
for blue and red galaxies.  The solid line is the C-M
(mass-metallicity) relation for E/S0's from Schombert \etal (1993).  Note
that the brightest galaxies are blue, and mostly galaxies with normal star
formation rates (Sp/Irr).}

\figcaption{Color-Magnitude diagram for the metallicity color, ($vz-yz$).
Photometric classifications are shown.  The solid line is the
C-M (mass-metallicity) relation for E/S0's from Schombert
\etal (1993).  There is little change in the mass-metallicity relation for
this cluster relative to the local population.}

\figcaption{The fraction of blue galaxies, $f_B$, as a function of
absolute magnitude.  The highest fraction of blue galaxies are also the
brightest members of the cluster. The rise at faint luminosities signals
the increasing contribution from low mass dwarfs and late-type Irr/Im/Sm
types.}

\figcaption{The number density of blue and red galaxies as a function of
cluster radius.  The red population dominates the core of the cluster with
the blue population having a marked decrease inside 0.3 Mpc.  This is in
agreement with one of the predictions of ``galaxy harassment" models,
where the blue population is due to tidally induced star formation from
the cluster and the most massive members.}

\clearpage
\begin{deluxetable}{lccc}
\small
\tablecolumns{10}
\tablewidth{8.0cm}
\tablecaption{A2317 cluster population}
\tablehead{\colhead{type} & \colhead{blue} & \colhead{red} & \colhead{total}}
\startdata
E/S0 & ~~0 (~0\%) & 47 (64\%) & 47 (42\%) \nl
Sp/Irr & 18 (46\%) & 11 (15\%) & 29 (26\%) \nl
Seyfert & 11 (29\%) & ~~0 (~0\%) & 11 (10\%) \nl
starburst & $\underline{\rm 10 ~(25\%)}$ & $\underline{\rm 15 ~(21\%)}$ & $\underline{\rm 25 ~(22\%)}$ \nl
& 39 (35\%) & 73 (65\%) & 112~~~~~~~~ \nl
\enddata
\end{deluxetable}

\pagestyle{empty} 
\clearpage
\begin{figure}
\plotfiddle{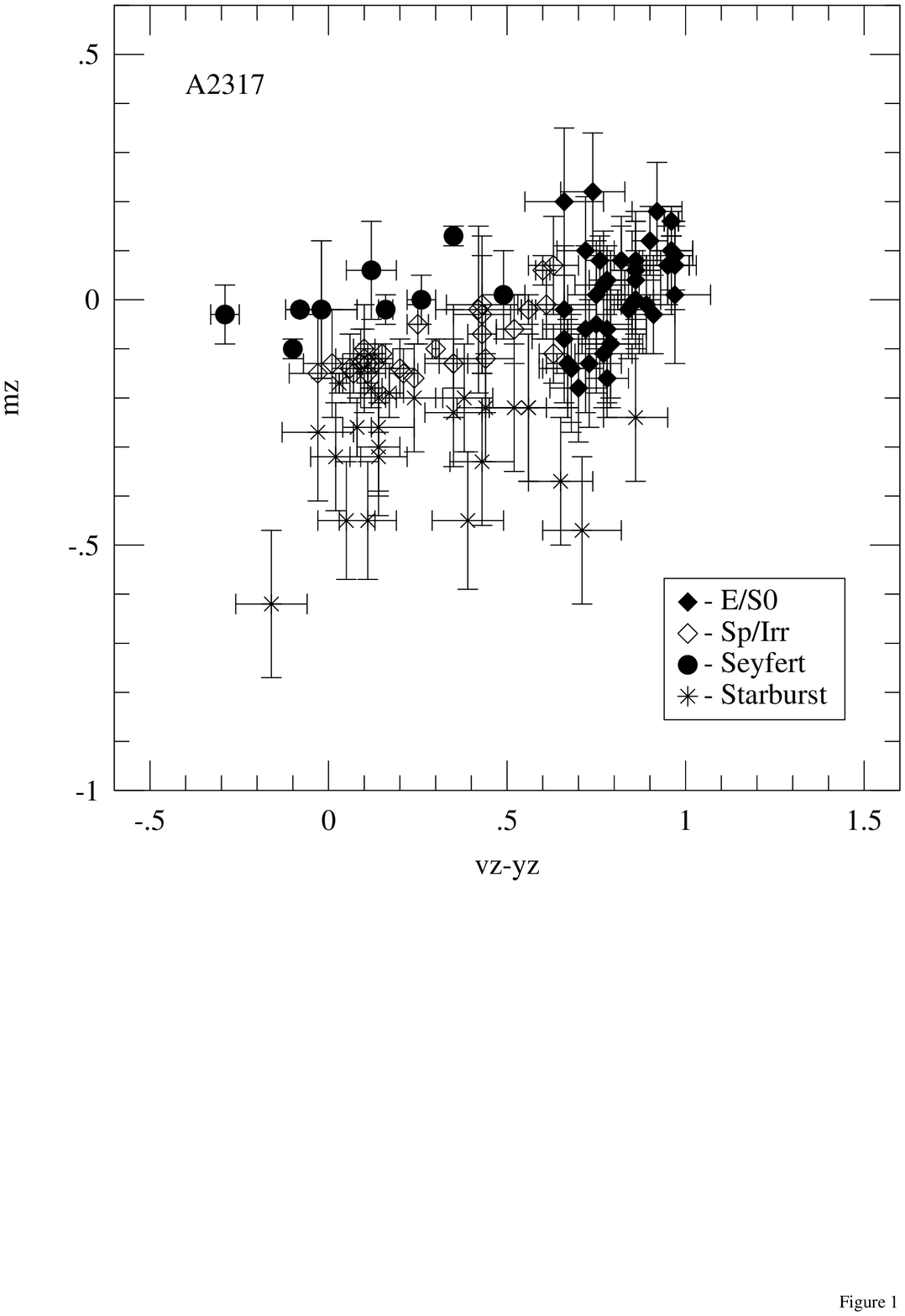}{11.5truein}{0}{100}{100}{-310}{170} \end{figure} 

\clearpage
\begin{figure}
\plotfiddle{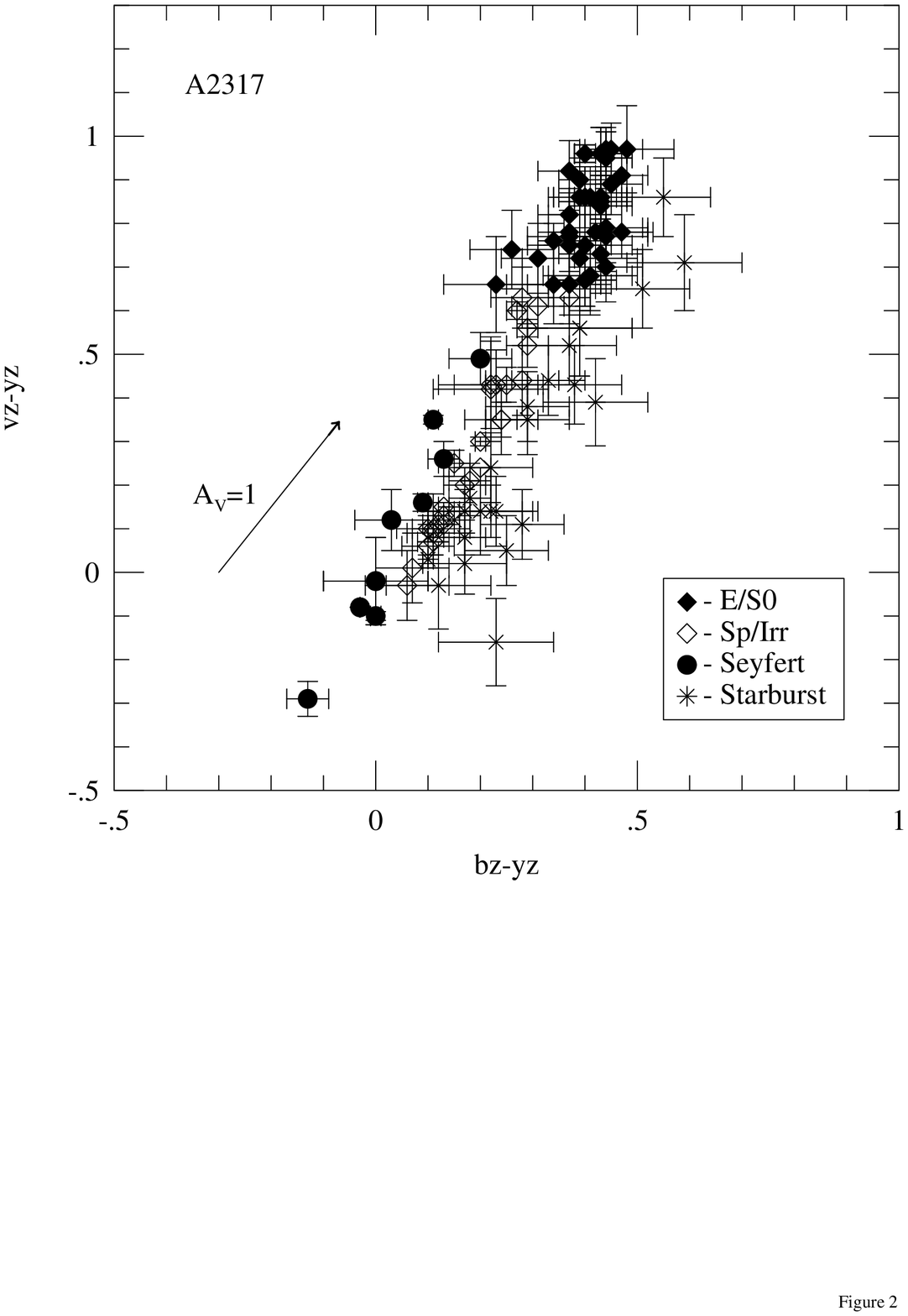}{11.5truein}{0}{100}{100}{-310}{170} \end{figure} 

\clearpage
\begin{figure}
\plotfiddle{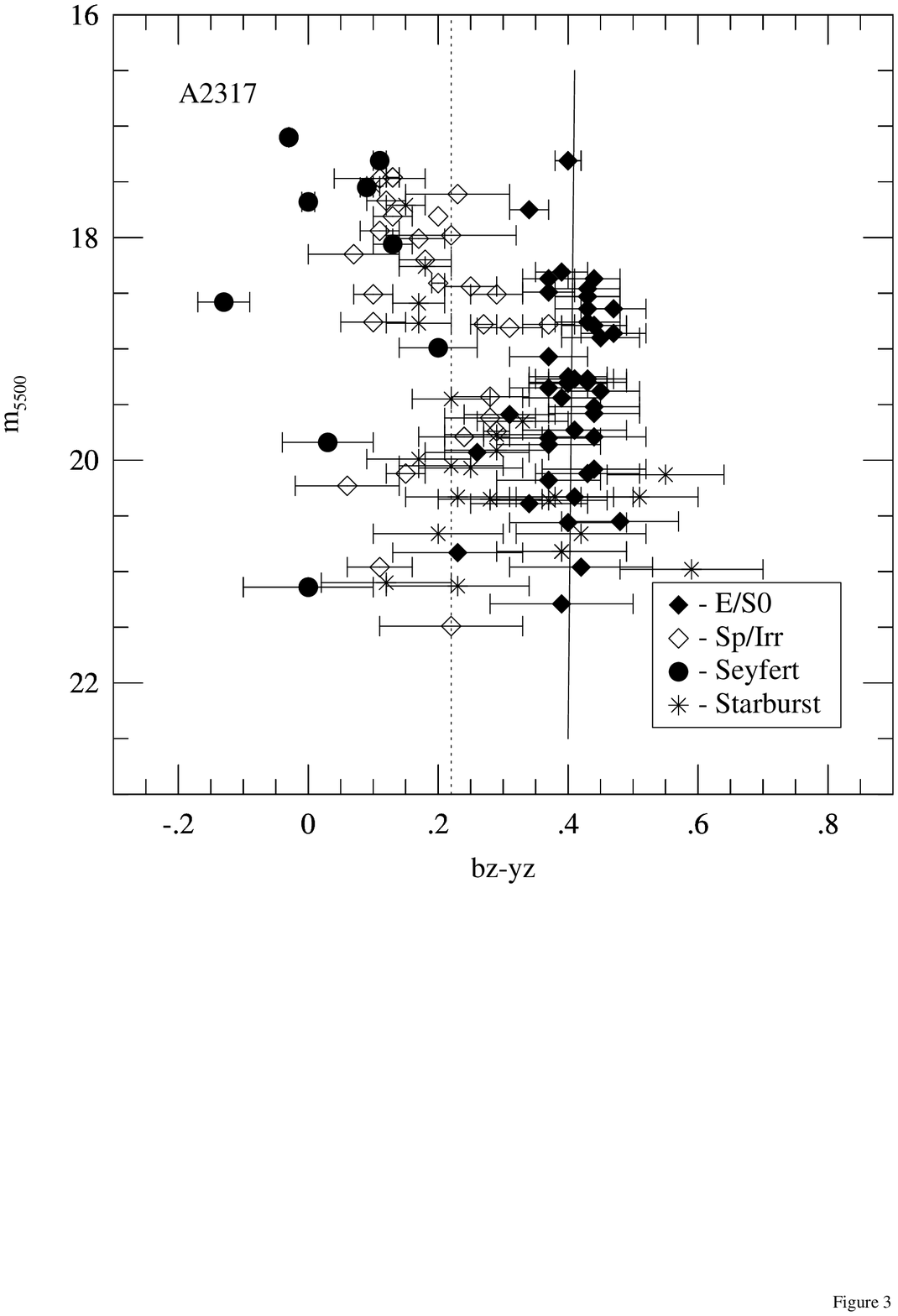}{11.5truein}{0}{100}{100}{-310}{170} \end{figure} 

\clearpage
\begin{figure}
\plotfiddle{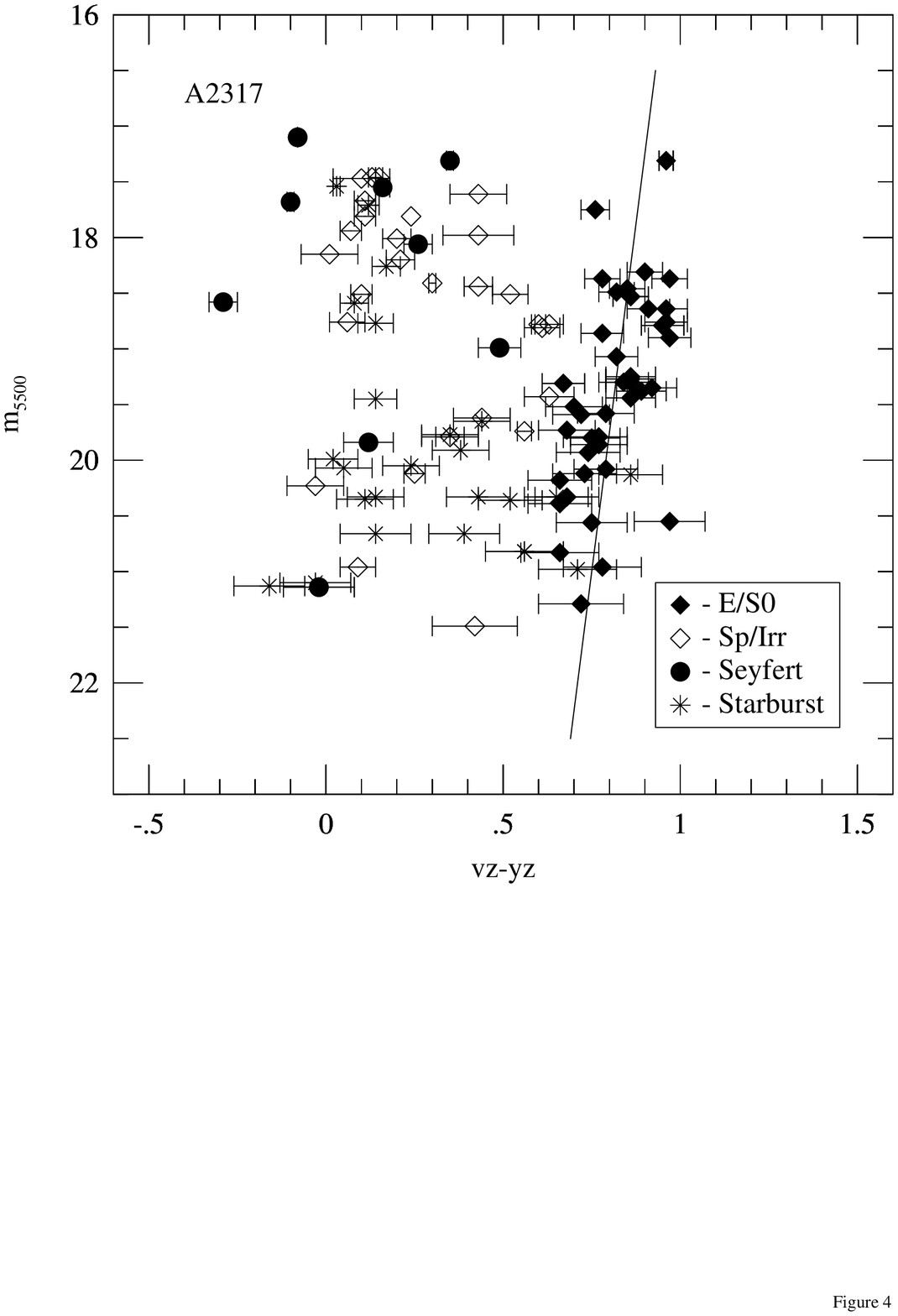}{11.5truein}{0}{100}{100}{-310}{170} \end{figure} 

\clearpage
\begin{figure}
\plotfiddle{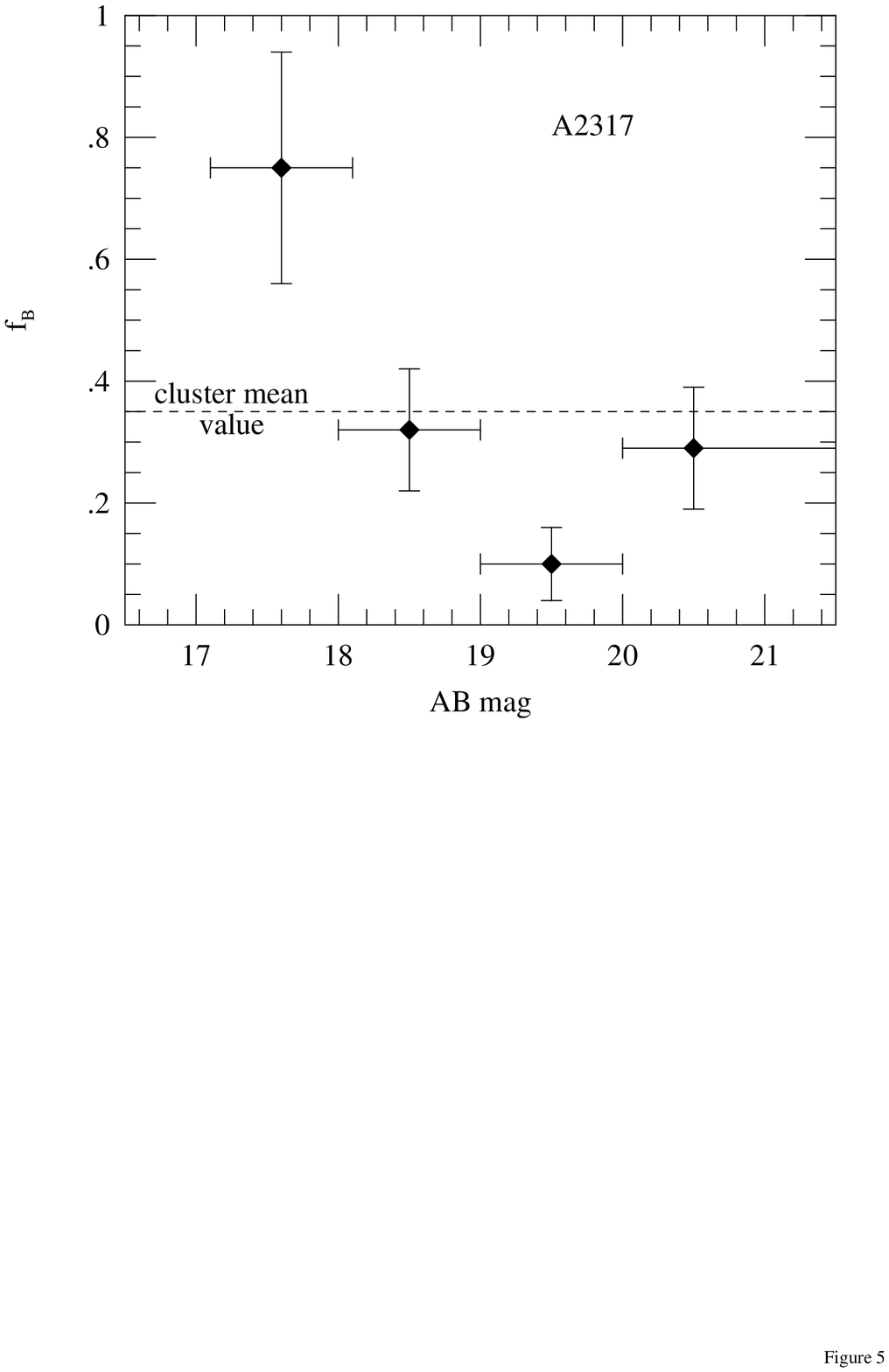}{11.5truein}{0}{100}{100}{-310}{170} \end{figure} 

\clearpage
\begin{figure}
\plotfiddle{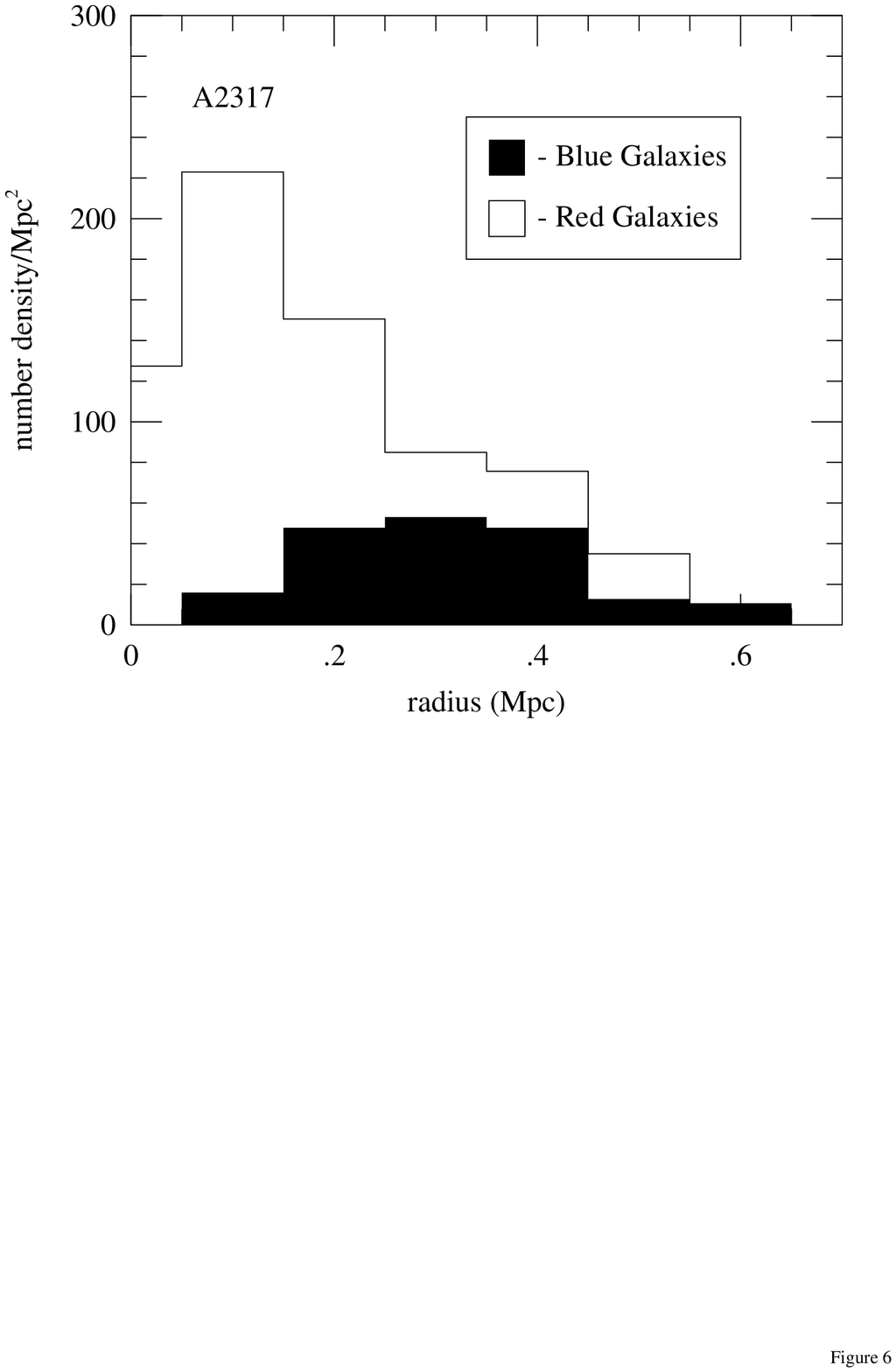}{11.5truein}{0}{100}{100}{-310}{170} \end{figure} 

\end{document}